# Sustainable and green synthesis of hydrogen tungsten bronze nanoparticles with nanocarbon via mechanically induced hydrogen spillover

Kunihiko Kato, [a] Takafumi Sudo, [b] Yunzi Xin, [a] and Takashi Shirai *[a]

a. Advanced Ceramics Research Center, Nagoya Institute of Technology, Gokiso, Showa-ku, Nagoya, Aichi 466-8555 Japan.

b. Department of Life Science and Applied Chemistry, Graduate School of Engineering, Nagoya Institute of Technology, Gokiso, Showa-ku, Nagoya, Aichi 466-8555 Japan.

## Abstract

In this study, we demonstrate sustainable and green nanotechnology for room-temperature synthesis of $H_xWO_3$ ($0<x<0.5$) via a novel reaction pathway induced by mechanical energy. A simple mixture of monoclinic $WO_3$ powder and polyolefin (polypropylene) is used to obtain $H_xWO_3$ nanoparticles that show high crystallinity even through high-energy ball milling synthesis. The composite of $H_xWO_3$ nanoparticles and nanocarbon by-products exhibit unique optoelectronic properties along with outstanding enhancement of photocatalytic performance in the decomposition of azo-dye water pollutants under visible light. The formation mechanism of the obtained functional material is also discussed. The findings of this study provide insights into the limitations for mass production of $H_xWO_3$ nanoparticles, such as a specific setup for electrochemical reactions and precious metal catalysis.

## 1. Introduction

Transition metal bronze (e.g., tungsten or molybdenum bronze) nanoparticles have attracted considerable attention owing to their high carrier density, high electrical conductivity [1], and intense light-absorbing capacity ranging from the visible to near-infrared (NIR) region [2], [3], revealed via localized surface plasmon resonance (LSPR), similar to metal nanoparticles (e.g., Au and Ag) [4], [5]. Thus, they are used in various applications such as superconductors [6], fuel cells [7], supercapacitor [8], smart windows [9], [10], and heterogeneous catalysts [11], [12]. These bronzes have a general formula of $M_xWO_3$ or $M_xMoO_3$, where M is usually $H^+$, alkali metal, alkaline earth metal, or lanthanide intercalated in the $WO_3$ or $MoO_3$ lattice. Typically, a high-temperature thermal treatment (at minimum of 500 K) is required to form $H_xWO_3$ via dissociative chemisorption of hydrogen on the surface of $WO_3$ as a hydrogen spillover (H-spillover) reaction [13], unless otherwise associated with precious metal catalysts (e.g., Pt or Pd) [12]. H-spillover is one of the main pathways for the formation of $H_xWO_3$ under thermochemical conditions, where $H_2$ or another hydrogen donor dissociatively adsorbs at the metal catalyst sites and the resultant surface-bound hydrogen migrates to the $WO_3$ support. An alternative approach is electrochemical formation through hydrogen intercalation in acidic electrolytes. The dynamics of $H_xWO_3$ bronze formation from $WO_3$ via thermochemical as well as electrochemical reduction were investigated using time-resolved optical microscopy [14], revealing that the reaction occurs via a congruent mechanism involving electron transfer coupled with proton uptake. Industrial mass production is limited by a specific setup for electrochemical reactions and precious metal catalysis in a hydrogen-rich atmosphere to overcome potential energy barriers [13]. Recently, some research groups have reported a mechanically induced reaction to synthesize $H_xWO_3$ as a third pathway. In these cases, specific conditions are needed in the ball milling process for efficient formation, such as the use of metastable $WO_3$ (hexagonal phase) [15] or metal catalyst [16], a reactive atmosphere (e.g., hydrogen) [15, 17].

In this study, we demonstrate a novel, sustainable, and green nanotechnology for the room-temperature synthesis of $H_xWO_3$ (0<x<0.5) via mechanically induced H-spillover in an inert atmosphere without utilizing precious metal catalysts, using a simple mixture of monoclinic $WO_3$ powder (m-$WO_3$) and polyolefin (polypropylene; PP). The proposed synthesis involves the simultaneous mechanical activation of $WO_3$, which facilitates the catalytic production of gaseous fragments as proton suppliers in an inert atmosphere, accompanied by the generation of graphitic nanocarbon as a by-product. The prepared $H_xWO_3$ has a highly ordered nanostructure and demonstrates outstanding enhancement of the photocatalytic reaction rate by an order of magnitude toward decomposing azo-dye water pollution under visible light.

## 2. Experimental

Commercially available highly pure $WO_3$ powder (>99.5%, Nacalai Tesque Co., Ltd.) and PP (Seishin Enterprise Co., Ltd.) powder were used as raw materials. These (powder volume ratio of $WO_3$ and PP = 6:1) were subjected to high-energy ball milling (rotation speed = 400 rpm for 3 h in an Ar-filled glovebox) in a commercial planetary ball mill apparatus (Pulverisette-7 premium line, Fritsch GmbH) using zirconia balls (diameter = 10 mm) and a zirconia vessel (80 mL).

The composition of the studied samples was investigated using X-ray diffraction (XRD, Ultima-IV, Rigaku). Raman spectroscopy was carried out using a Raman spectrometer (NRS-3100, Jasco, or Via Raman microscope, Renishaw plc.). Fourier transform infrared (FT-IR) spectra were recorded using an FT-IR spectrometer (FT-IR 6600, JASCO). Nanostructured surfaces of the as-synthesized particles were examined using transmission electron microscopy (TEM; JEM-2100F, JEOL). Particle morphology was analyzed using field-emission scanning electron microscopy (JSM-7000F, JEOL). The chemical state of the initial sample surface was investigated using X-ray photoelectron spectroscopy (XPS, M-probe, Surface Science Instrument) with an Al Kα source (hν = 1486.6 eV). Electron spin resonance (ESR) spectra were measured using a JEOL JES-FA200 spectrometer. Optical properties in the UV to visible range and the UV to the near-infrared range were analyzed using UV-vis spectrophotometry (V-7100, Jasco).

Photocatalytic performance was evaluated by the degradation of methyl orange (MO) as an azo-dye water pollutant under visible light. First, 5 mg of either raw $WO_3$ or the as-prepared powder were loaded into a vessel containing 10 mL of an MO solution at 15 ppm. A 300 W Xe lamp (MAX-350, ASAHI SPECTRA, Japan) with a band-path filter (385–740 nm) was used as the visible-light source. Before irradiation, the solution containing the $WO_3$ photocatalysts was allowed to sit in the dark for approximately 60 min to ensure equilibrium adsorption of the dye molecules. Thereafter, MO degradation was monitored by measuring changes in the UV-vis absorption spectra at regular intervals under light irradiation. Photoluminescence (PL) spectra were measured using a commercial spectrofluorometer (FP-8500, Jasco, Japan) at an excitation wavelength of 350 nm.

## 3. Results and discussion

The XRD patterns show the phase transformation of $WO_3$ from monoclinic to tetragonal in the as-prepared powder only through milling with PP (S2) (**Fig. 1**), whereas crystallinity decreases due to the strong lattice distortion without PP during the high-energy ball milling process (S1). Considering that $M_xWO_3$ exhibits a tetragonal phase ($0<x<0.5$) [15], the obtained material with a tetragonal phase was identified as $H_xWO_3$. The presence of tetragonal $H_xWO_3$ with a highly ordered nanostructure was confirmed by TEM (**Fig. 2a**). Although the mechanically treated powder shows lattice distortion (black contrast in Fig. 2b), amorphous layers (thickness of few nanometers) were formed initially on the surfaces of the particles, which are typically found in ceramic powders after the milling process [18]. In contrast, the ball-milled $WO_3$ nanoparticles without PP (S1) display the aggregation and the amorphous surface structure (presented in Fig. 2b). Raman spectra shown in **Fig. 3a** show that the spectrum of the m-$WO_3$ raw powder displays two intense peaks at 720 cm$^{-1}$ and 807 cm$^{-1}$, which are attributed to phonons in crystalline $WO_3$ [19]. In the mechanically treated $WO_3$ powder (S1), a weak peak appeared at 940 cm$^{-1}$, which was assigned to the amorphous phase of $WO_3$ [19]. In contrast, ball milling with PP produces graphitic nanocarbon as a by-product with intense G-band and D-band peaks (**Fig. 3b**). These peaks represent in-plane $sp^2$ bonded carbons in an aromatic ring stretching and disorder in the $sp^2$ carbon network, respectively [20]. The by-product nanocarbon plays an essential role in protecting the generated $H_xWO_3$ lattice from further breakdown and amorphization during high-energy ball milling. As seen from the XPS spectra of the W4f orbital (**Fig. 4a**), the broadened peak toward lower binding energy attributed to $W^{5+}$ species [21] appears in the treated $WO_3$ (S1) compared to the pristine powder (S0). As reviewed in Ref [22], the mechanical stressing during ball milling initiates the formation of defects such as oxygen vacancies, causing the generation of $W^{5+}$ species. Besides, the synthesized $H_xWO_3$ (S2) also displays $W^{5+}$ species. It is well known that migration of proton and electron proceeds in the bulk of $WO_3$ lattice via classical H-spillover, leading to the formation of $H_xWO_3$ with $W^{5+}$ species [8,23]. The FT-IR spectra exhibit further evidence of the $H_xWO_3$ formation (**Fig. 4b**). The broadband in the range of 850–650 cm$^{-1}$ is attributed to the W-O stretching vibrations of crystalline

WO$_3$ [23]. An increase in wide-ranging IR absorption above 1100 cm$^{-1}$ was observed in the prepared powder through ball milling with PP (S2), reflecting the presence of quasi-free electrons that cause metallic reflectivity in H$_x$WO$_3$ [23]. An extra broad feature (<600 cm$^{-1}$) attributed to the W-O stretching modes strongly coupled with protons was also observed [23]. In this case, the C-H bond as residual organic carbon in the range of 2900–2700 cm$^{-1}$ disappeared in S2. Furthermore, the ESR spectra suggest a notable difference in the structure formed through the ball milling with polyolefin. As shown in **Fig. 4c**, S1 displays a strong signal at g = 2.004 associated with unpaired electrons trapped by oxygen vacancies [24, 25] rather than by W$^{5+}$ (g = 1.89 [24], 1.94 or 2.19 [25]). In contrast, the intense radical in S2 can be assigned as not the trapped electrons in oxygen vacancies but free-electrons in H$_x$WO$_3$, which is consistent with the broad IR adsorption above 1100 cm$^{-1}$.

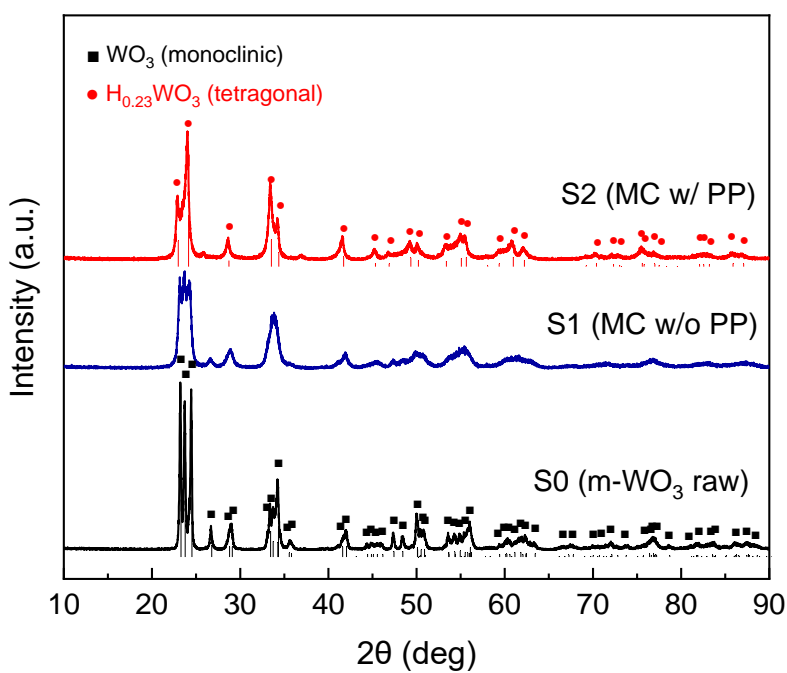

**Fig. 1.** XRD patterns of the raw and the pristine m-WO$_3$ powders after the ball milling.

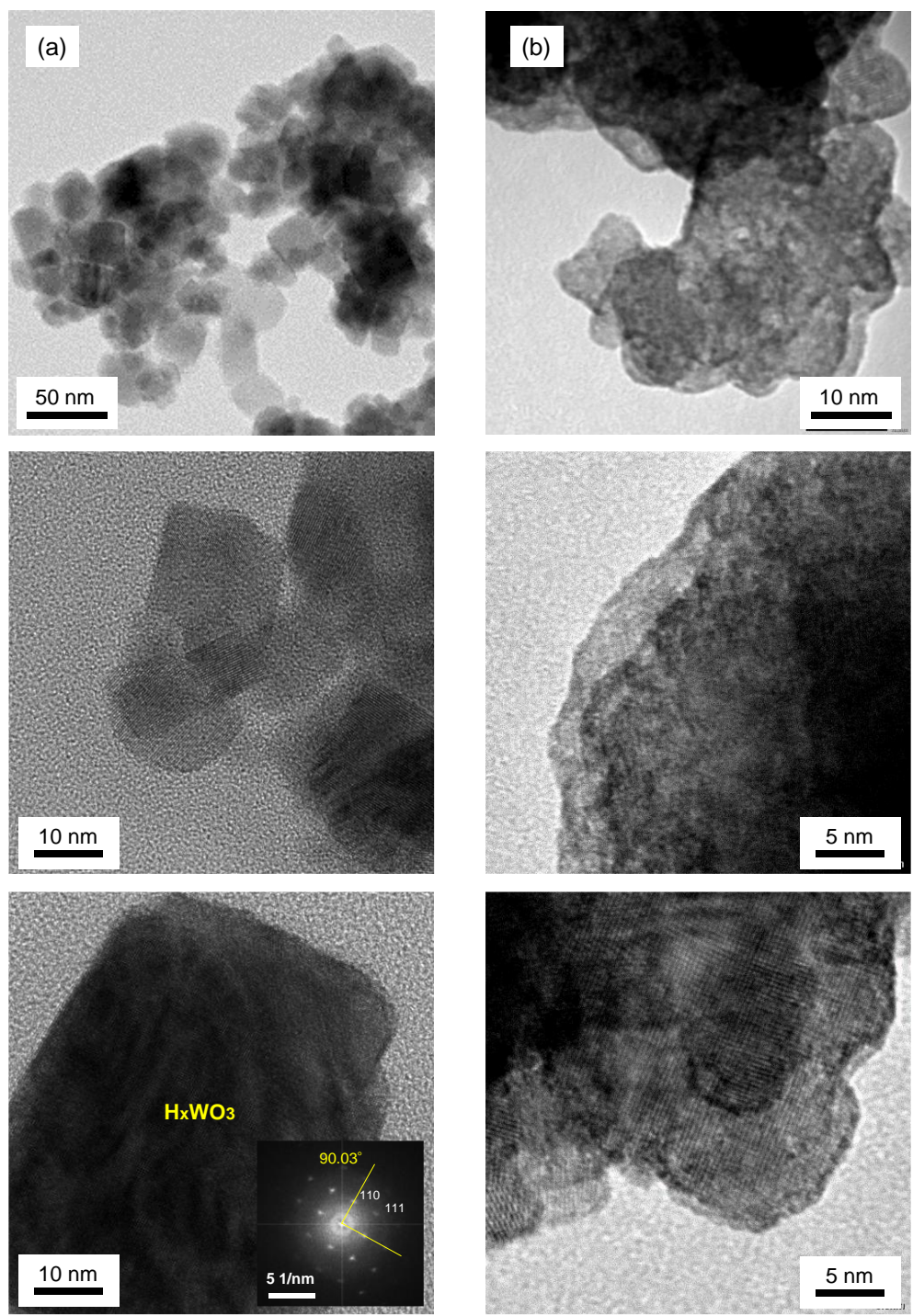

**Fig. 2.** TEM images; (a) the $H_xWO_3$ nanoparticles (sample "S2", left side), (b) the ball-milled $WO_3$ nanoparticles without PP (sample "S1", right side).

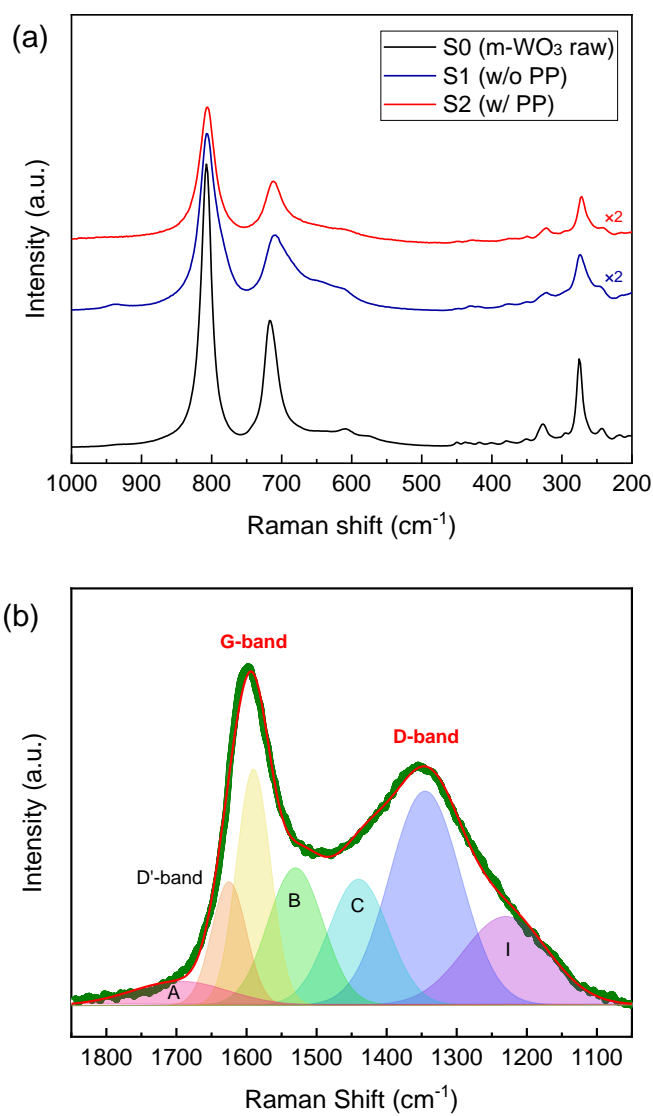

**Fig. 3.** Raman spectra in the range from (a) 1000–100 cm$^{-1}$ and (b) 1080–1050 cm$^{-1}$.

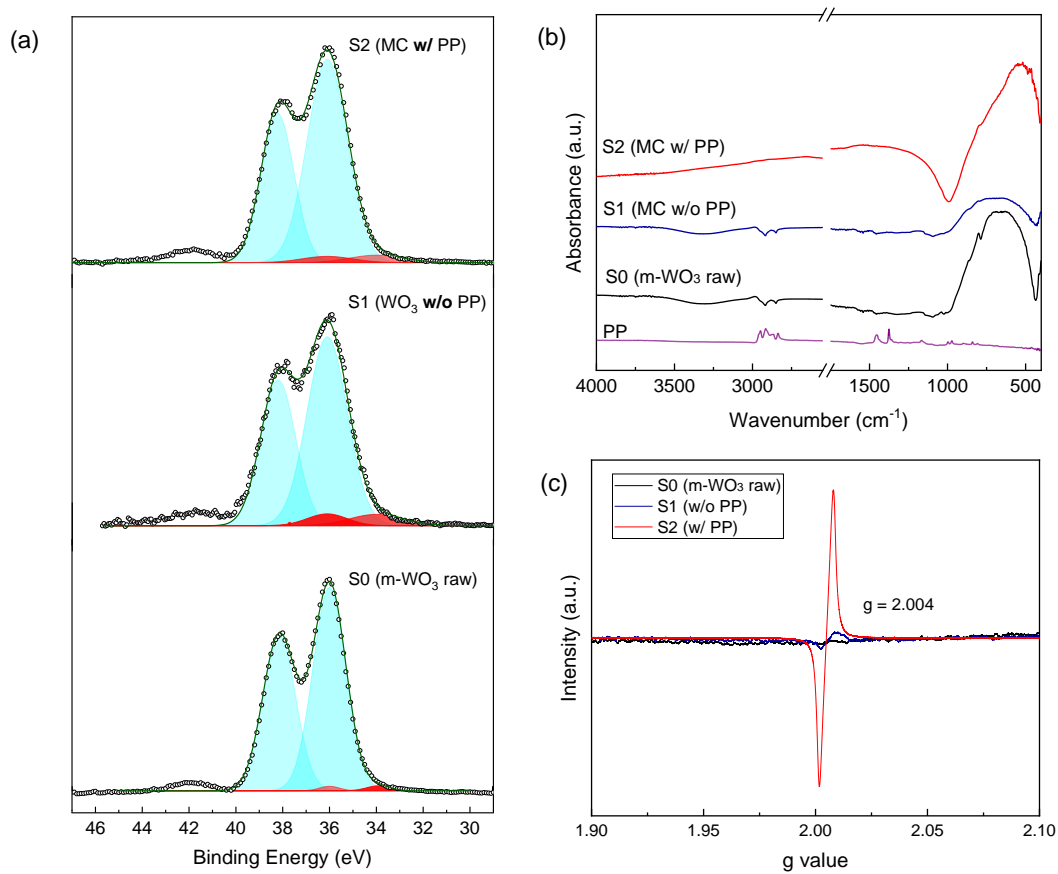

**Fig. 4.** (a) XPS spectra ($W_{4f}$ orbital), (b) FT-IR spectra, and (c) ESR spectra.

A significant increase in the light-absorbing capacity, ranging from the near-infrared region, was observed in the composite powder (**Fig. 5a-b**). A color change from yellow to blue associated with the formation of $H_xWO_3$ has been used to confirm the existence of atomic hydrogen via hydrogen spillover [13], [26]. It is due to the Fermi level shifts toward the CBM [27] because electrons are supplied by heavy hydrogen doping, giving a characteristic metallic feature such as LSPR in hydrogen metal bronze. This specific characteristic is featured only by tungsten and molybdenum oxides [3]. Herein, we need to note that S2 shows the red-shift in the absorption edge. A similar red-shift has been reported in semiconductor particle/nanocarbon composite, caused by the formation of the chemical bonding in the interface [28], [29]. In this case, it is difficult to determine the actual optical bandgap from UV-vis spectra since the presence of carbon component involves a continuous absorption band in the range of 400-800 nm.

The outstanding enhancement of photocatalytic activity is demonstrated by the photodegradation of MO under visible-light irradiation (>385 nm) compared to the pristine WO$_3$ and the mechanically treated WO$_3$ powders, as shown in **Fig. 6a-b**. MO possesses strong absorption at 465 nm and 270 nm, corresponding to the azo bond (-N=N-) and the benzene ring, respectively [30], [31]. The central absorption peak (465 nm) red-shifted and decreased quickly due to the cleavage of the conjugated double bond with decolorization, starting after photodegradation. Azo-dye is commonly used as a pH indicator that exhibits a color change from yellow to red with increasing acidity. The OH groups formed in H$_x$WO$_3$ would have Bronsted acidic features, functioning as active sites for the catalytic reaction in this case. Moreover, the absorption peak attributed to the benzene ring is shifted toward shorter wavelengths (240–260 nm) during photocatalysis, indicating that the degraded products could be attributed to sulfanilic acid and N, N-dimethyl-p-phenylenediamine [30], [31]. The new absorption of nitrophenols arises from the π (benzene ring)–π* (nitro group) transition at 330 nm [32], as well as the peak of amine [25] in the UV region (<240 nm) throughout photodegradation. The photocatalytic degradation kinetics follow the first-order law, expressed as ln(C$_0$/C) = $k$t, where t is the irradiation time, and $k$ is the degradation rate constant. The value of $k$ was determined as 2.79×10$^{-2}$ min$^{-1}$ from the slope of the linear graphs in **Fig. 6b**, which enhances the photocatalytic reaction rate by one order of magnitude. It is generally known that the H$_x$WO$_3$ phase possesses high electrical conductivity and promotes the charge separation of photoexcited carriers (electrons and holes) by forming a heterojunction with WO$_3$, leading to the enhancement of photocatalytic activity [11]. However, the observed visible-light photocatalytic activity of a single phase of $H_xWO_3$ is atypical. In evaluating the carrier recombination rates by photoluminescence (PL), a further decrease in PL intensity was observed in the synthesized composites (**Fig. 6c**). As a further experiment to understand the enhancement of photocatalytic activity in the synthesized material, the heat treatment was conducted for S2 at 400 °C (10 K/min without temperature keeping) in the inert atmosphere (Ar) to release hydrogen in $H_xWO_3$ [15], [17]. **Fig. 7a** shows the XRD patterns, where the phase transition from tetragonal ($H_xWO_3$) to monoclinic ($WO_3$) is found in the thermally treated particle (S3). As well as the characteristic IR absorption originated from $H_xWO_3$, this result supports the formation of $H_xWO_3$ via the ball-milling process. Moreover, S3 shows a thermal inactivation in the photocatalytic performance (displayed in **Fig. 7b**). Thus, the presence of $H_xWO_3$ should be a crucial factor in enhancing photocatalytic activity. In this case, the adsorption amount of methyl orange after adsorption/desorption equilibrium (reaction time = 0 min) is negligibly minor than the decomposition amount. Besides, the disappearance of $H_xWO_3$ causes a higher carrier recombination rate (shown in **Fig.7 (c)**). Nevertheless, the possibility should be noted that nanocarbon facilitates photocatalytic reaction as cocatalyst [33] more or less for efficient adsorption of dye molecules and excited carrier separation.

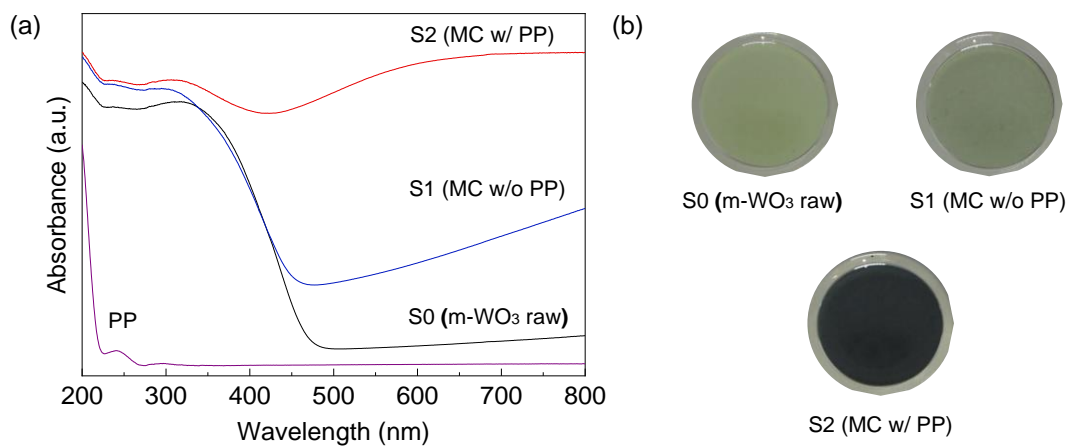

**Fig. 5.** (a) UV-vis spectra, (b) photographs of raw and prepared powders.

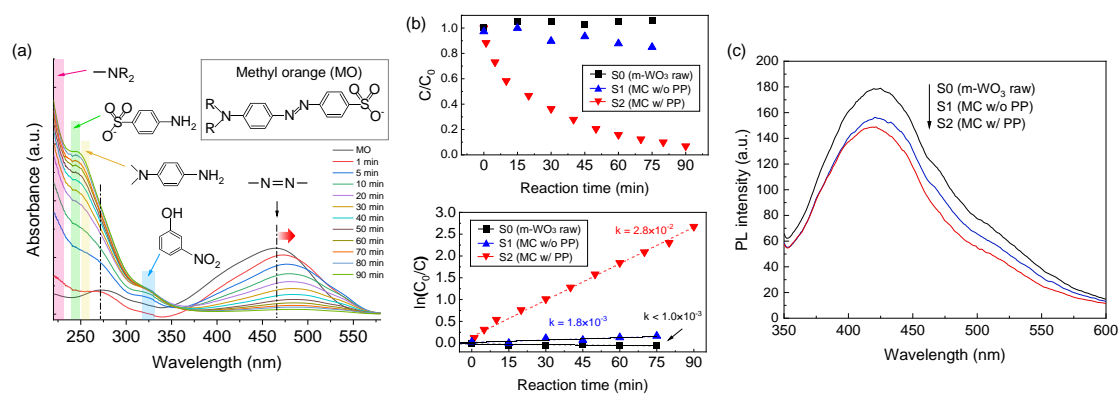

**Fig. 6.** (a) photocatalytic activity for the degradation of MO by S2 (synthesis with PP), (b) concentration change during visible-light irradiation (>385 nm), (f) photocatalytic reaction rate, and (c) PL spectra.

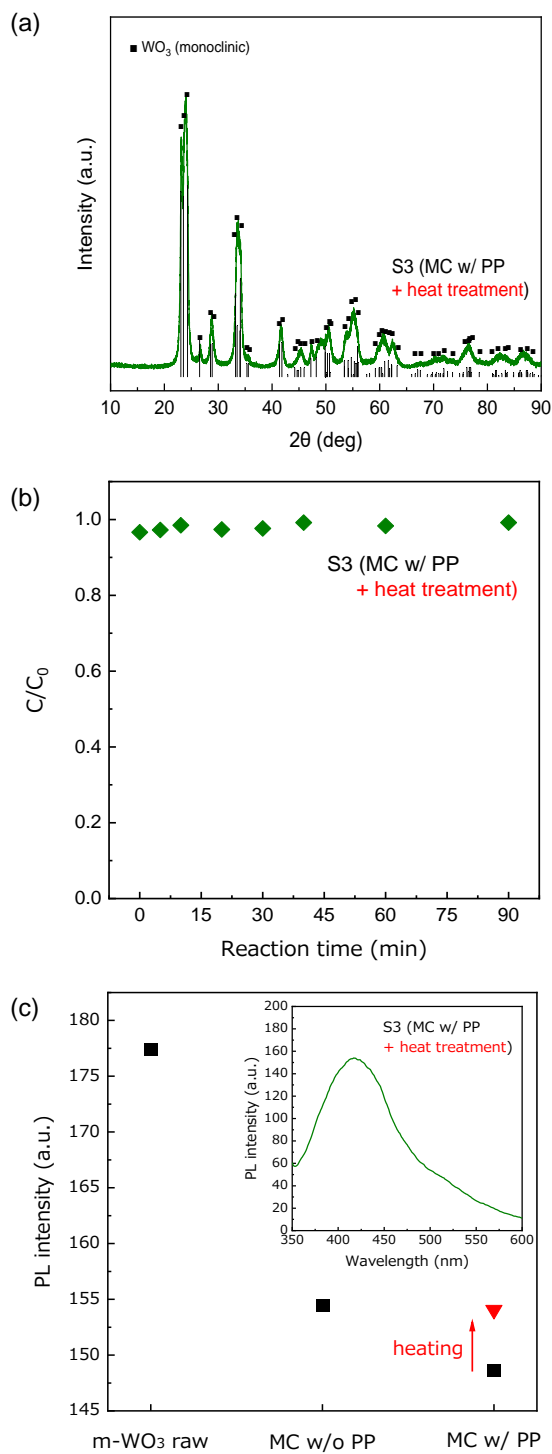

**Fig. 7.** Evaluation of the thermally treated sample; (a) XRD pattern, (b) the photocatalytic performance, and (c) comparison of PL intensity (the inset indicates PL specetra of S3).

To understand the specific reaction pathway for the room-temperature formation of $H_xWO_3$ without thermal or electrochemical effects, the in situ measurement of pressure change in the zirconia vessel was carried out during the mechanical treatment in various atmospheres (**Fig. 8a**). In the milling of m-WO$_3$ only, the pressure increased followed by saturation within 60 min, which might be due to the vaporization of physisorbed water molecules on the particle surface. In contrast, the milling of the mixture powder caused a considerable increase in pressure, resulting from the mechanical decomposition of polyolefin and leading to the production of the gaseous component of hydrocarbon fragments, similar to that reported in Ref [34]. Notably, a sharp decrease in pressure was observed in air condition followed by a temporary increase. It follows from the XRD patterns of the composite powder prepared in different atmospheres during the ball milling, that the phase transition from monoclinic to tetragonal is retarded greatly by the treatment in the air (**Fig. 8b**). Furthermore, a distinct change between the two different powders was observed in the structure of the carbonaceous by-product material from the Raman spectra (**Fig. 8c**). Mechanical treatment in an inert atmosphere causes a much more highly ordered graphitic structure, whereas the defective nanocarbon is preferably formed by ball milling in air. This fact suggests that the loss of active fragments acting as proton suppliers through oxidation negatively impacts the effective formation of $H_xWO_3$. Moreover, the structural difference is reflected by the optical property, and the powder treated in the oxidation atmosphere displays a weaker absorbance up to the visible region (>400 nm) (see **Fig. 8d**). As our co-authors have reported for a mechano-chemical reduction of metal oxides through high-energy ball milling in the presence of an organic polymer [35], this phenomenon might be associated with the reduction of the WO$_3$ surface in air. It follows from the XRD patterns shown in **Fig. 8e** that the mixed crystalline structures of m-WO$_3$ and tetragonal $H_xWO_3$ appeared after 1 h of treatment, and the phase transformation completed within 3–5 h during ball milling. The migration of hydrogen from the terminal oxygen on the molybdenum oxide surface to the bulk is known to be the rate-limiting step with an activation barrier of 0.6 eV [13]. Because the proton diffusion in the solid-state of $WO_3$ is much faster (50–175 μm/s) [14] than the observed time lag to complete the phase transformation, the structural modifications such as lattice distortion have to be a key component to serve as the catalytic function in H-spillover promoted by the active fragment to lower the energy potential.

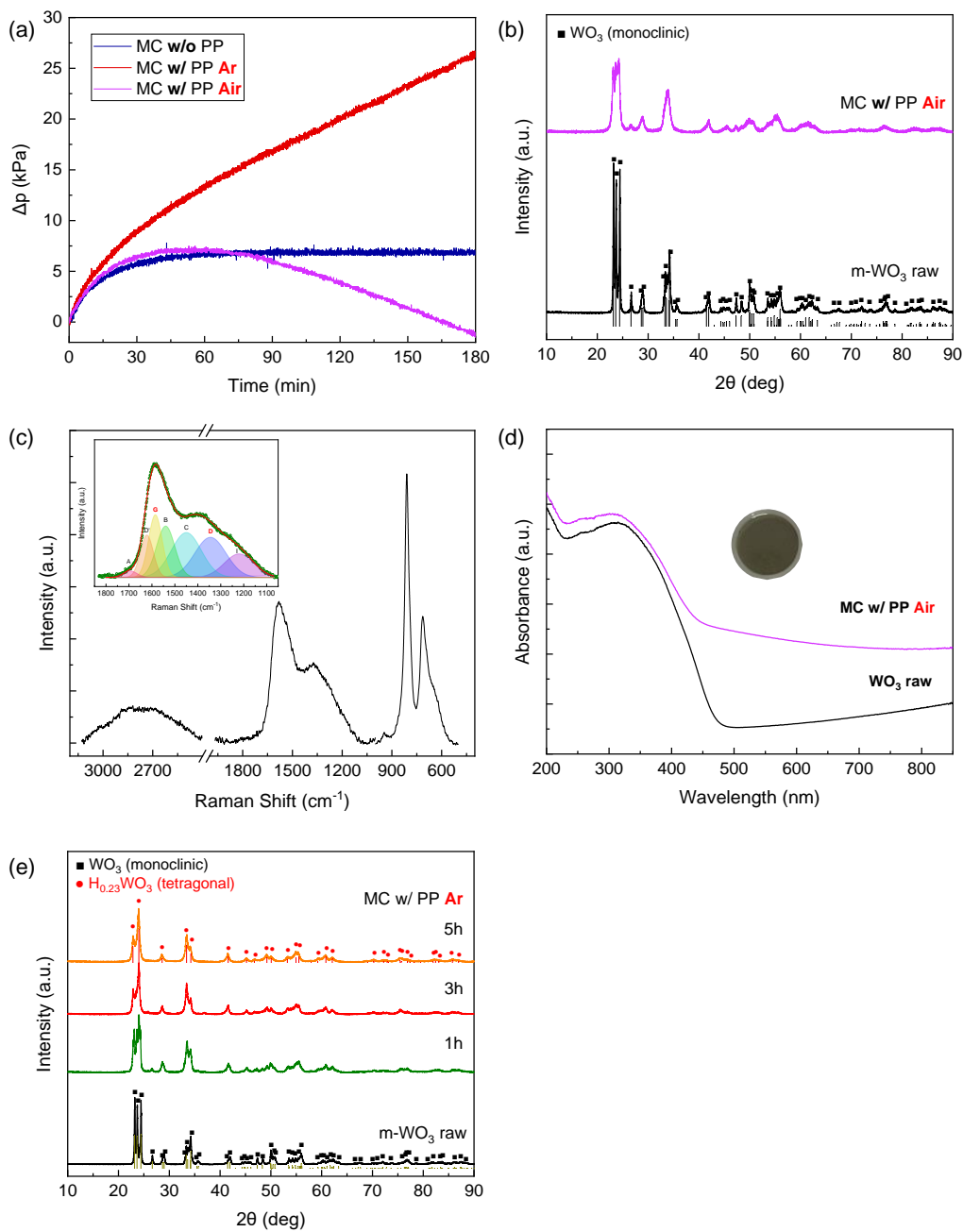

**Fig. 8.** Comparison of synthesis with the different atmospheres (Ar vs. Air); (a) in situ measurement of pressure change in the zirconia vessel during ball milling, (b) XRD patterns, (c) Raman spectra, (d) UV-vis spectra. (e) XRD patterns of prepared samples with varied treatment time.

In this study, we propose a new scheme for the room-temperature formation of $H_xWO_3$ via mechanically induced H-spillover without precious metal catalysis (**Fig. 9**). The high-energy ball milling process is associated with the simultaneous lattice distortion of m-$WO_3$ and the mechanical depolymerization of PP in the first stage. In an inert atmosphere, the generation of a gaseous fragment is effectively facilitated, as the defective surface of the transition metal oxide assists the catalytic reaction to further decomposition of hydrocarbons. It must be a pivotal pathway for room-temperature H-spillover to form $H_xWO_3$ (tetragonal) with a highly ordered nanostructure. On the other hand, the strongly chemisorbed intermediate carbon species efficiently converted into the graphitic nanocarbon as a by-product, protecting the $H_xWO_3$ lattice from further breakdown and amorphization. However, the formation of $H_xWO_3$ is firmly blocked in an air atmosphere because of the physical/chemical stabilization or relaxation of active fragments and the modified $WO_3$ surface by oxidation.

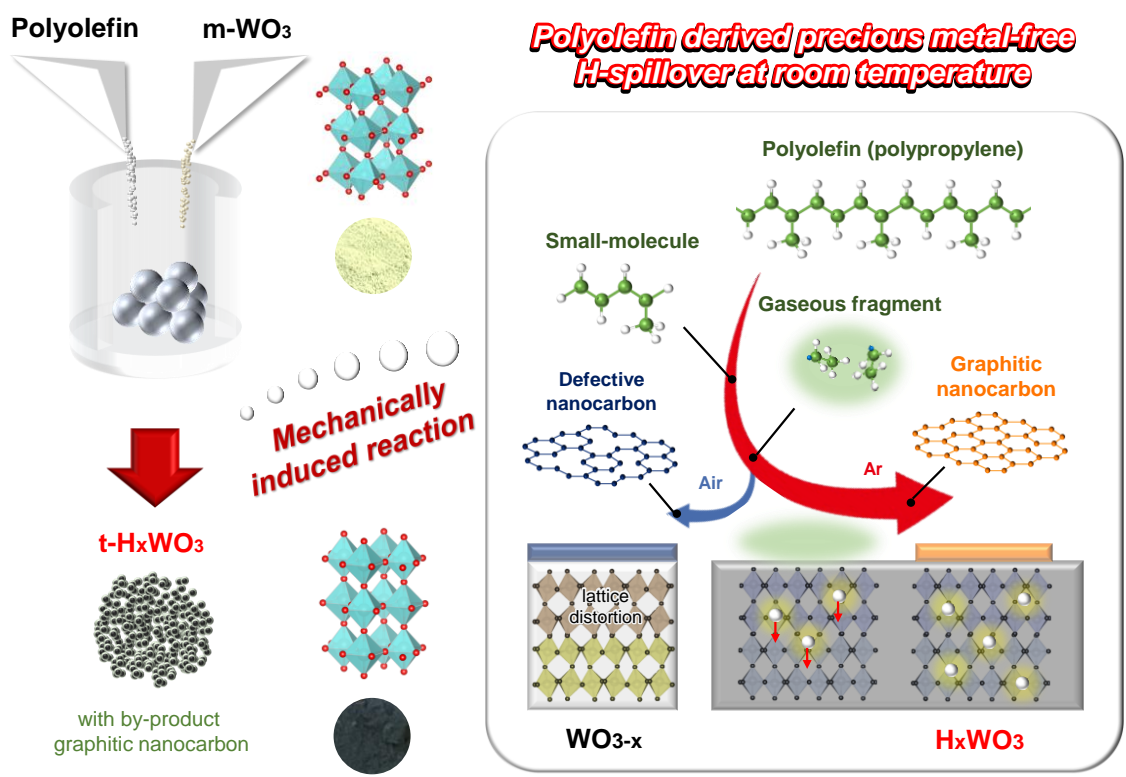

**Fig. 9.** Schematic illustration of the room-temperature synthesis of $H_xWO_3$ in mechanically induced H-spillover.

## 4. Conclusions

We proposed the room-temperature synthesis of $H_xWO_3$ achieved by a mechanically induced reaction under inert atmospheric conditions, where the starting materials include only a dry powder mixture of m-$WO_3$ and PP. The synthesized highly crystalline $H_xWO_3$ nanoparticles (tetragonal; 0<x<0.5) showed a specific optoelectronic feature, exhibiting outstanding enhancement of photocatalytic activity toward the degradation of azo-dye water pollutants (MO) under visible light. Furthermore, the formation mechanism of $H_xWO_3$ with the graphitic nanocarbon by-product is also addressed, involving H-spillover on the mechanically activated m-$WO_3$ surface. The catalytic generation of a gaseous fragment as a proton supplier would be effectively facilitated in an inert atmosphere followed by the simultaneous mechanical depolymerization of PP. Additional work is required for detailed understanding of the mechanism and electronic structure of $H_xWO_3$-nanocarbon composites, however, we believe that the sustainable and green nanotechnology for the synthesis of $H_xWO_3$ breaks down the limitation for mass production constrained from the use of a specific setup for electrochemical reactions and precious metal catalysis to surmount the potential energy barrier for H-spillover.

# References

[1] T.M. Mattox, A. Bergerud, A. Agrawal, D.J. Milliron, Influence of Shape on the Surface Plasmon Resonance of Tungsten Bronze Nanocrystals, Chem. Mater. 26 (2014) 1779–1784.

[2] C. Yang, J.F. Chen, X. Zeng, D. Cheng, D. Cao, Design of the alkali-Metal-Doped $WO_3$ as a Near-Infrared Shielding Material for Smart Window, Ind. Eng. Chem. Res. 53 (2014) 17981–17988.

[3] H. Cheng, M. Wen, X. Ma, Y. Kumahara, K. Mori, Y. Dai, B. Huang, H. Yamashita, Hydrogen Doped Metal Oxide Semiconductors with Exceptional and Tunable Localized Surface Plasmon Resonances, J. Am. Chem. Soc., 138 (2016) 9316.

[4] C.L. Haynes, R.P.V. Duyne, Nanosphere Lithography:  A Versatile Nanofabrication Tool for Studies of Size-Dependent Nanoparticle Optics, J. Phys. Chem. B 105 (2001) 5599–5611.

[5] A.J. Haes, R.P.V. Duyne, A Nanoscale Optical Biosensor:  Sensitivity and Selectivity of an Approach Based on the Localized Surface Plasmon Resonance Spectroscopy of Triangular Silver Nanoparticles, J. Am. Chem. Soc. 124 (2002) 10596–10604.

[6] S. Reich, G. Leitus, R.P. Biro, A. Goldbourt, S. Vega, A Possible 2D $H_xWO_3$ Superconductor with a Tc of 120 K, J. Supercon. Nov. Mag. 22 (2009) 343–346.

[7] D. Yoon, A. Manthiram, Hydrogen tungsten bronze as a decoking agent for long-life, natural gas-fueled solid oxide fuel cells, Energ. Environ. Sci. 7 (2014) 3069–3076.

[8] H. Wang, R. Fan, J. Miao, J. Chen, S. Mao, J. Deng, Y. Wang, , J. Mater. Chem. A 6 (2018) 6780–6784.

[9] Y. Lee, T. Lee, W. Jang, A. Soon, Unraveling the Intercalation Chemistry of Hexagonal Tungsten Bronze and Its Optical Responses, Chem. Mater. 28 (2016) 4528–4535.

[10] P.J. Wu, S. Brahma, H.H. Lu, J.L. Huang, Synthesis of cesium tungsten bronze by a solution-based chemical route and the NIR shielding properties of cesium tungsten bronze thin films, Appl. Physica A 126 (2020) 98.

[11] L. Zhang, W. Wang, S. Sun, D. Jiang, Near-infrared light photocatalysis with metallic-semiconducting $H_xWO_3$-$WO_3$ nanogeterostructure in situ formed in mesoporous template, Appl. Catal. B 168 (2015) 9–13.

[12] Y.F. Li, N. Soheilnia, M. Greiner, U. Ulmer, T. Wood, A.A. Jelle, Y. Dong, A.P.Y. Wong, J. Jia, G.A. Ozin, Pd@$H_yWO_{3-x}$ Nanowires Efficiently Catalyze the $CO_2$ Heterogeneous Reduction Reaction with a Pronounced Light Effect, ACS Appl. Mater. Interfaces 11 (2019) 5610–5615.

[13] Y. Xi, Q. Zhang, H. Cheng, Mechanism of Hydrogen Spillover on $WO_3$(001) and Formation of $H_xWO_3$ (x = 0.125, 0.25, 0.375, and 0.5), J. Phys. Chem. C 118 (2014) 494–501.

[14] E.V. Miu, J.R. McKone, Comparison of $WO_3$ reduction to $H_xWO_3$ under thermochemical and electrochemical control, J. Mater. Chem. A, 7 (2019) 23756–23761.

[15] F.J. Castro, F. Tonus, J.L. Bobet, G. Urretavizcaya, Synthesis of hydrogen tungsten bronzes $H_xWO_3$ by reactive mechanical milling of hexagonal $WO_3$, J. Alloy. Compd. 495 (2010) 537–540.

[16] Y. Cui, F. Liang, C. Ji, S. Xu, H. Wang, Z. Lin, J. Liu, Discoloration Effect and One-Step Synthesis of Hydrogen Tungsten and Molybdenum Bronze ($H_xMO_3$) using Liquid Metal at Room Temperature, ACS Omega 4 (2019) 7428–7435.

[17] G. Urrettavizcaya, F. Tonus, EE. Gaudin, J.L. Bobet, F.J. Castro, Formation of tetragonal hydrogen tungsten bronze by reactive mechanical alloying, J. Solid State Chem. 180 (2007) 2785–2789.

[18] S. Takai N. Hoshimi, T. Esaka, Synthesis of Tungsten, Molybdenum and Vamdium Bronzes by Mechanochemical Method Milling with Liquid Hydrocarbons, Electrochemistly, 72 (2004) 876–879.

[19] S. Indris, R. Amade, P. Heitjans, M. Finger, A. Haeger, D. Hesse, W. Frunert, A. Borger, K. D. Becker, Preparation by High-Energy Milling, Characterization, and Catalytic Properties of Nanocrystalline $TiO_2$, J. Phys. Chem. B, 109 (2005) 23274–23278.

[20] S.H. Lee, H.N. Cheong, C.E. Tracy, A. Mascarenhas, D.K. Bensom, S.K. Deb, Raman spectroscopic studies of electrochromic a-$WO_3$, Electrochimica Acta 44 (1999) 3111–3115.

[21] L. Bokobza, J.L. Bruneel, M. Couzi, Raman spectroscopic investigation of carbon-based materials and their composites. Comparison between carbon nanotubes and carbon black, *Chem. Phys. Lett.*, 590 (2013) 153–159.

[22] P. Balaz M. Achimovicova, M. Balaz, P. Billik, Z.C. Zheleva, J.M. Criado, F. Delogu, E. Dutkova, E. Gaffet, F.J. Gotor, R. Kumar, I. Mitov, T. Rojac, M. Senna, A. Streletskii, K.W. Ciurowa, Hallmarks of Mechanochemistry: From Nanoparticles to Technology, Chem. Soc. Rev. 42 (2013) 7571-7673.

[23] H. Tian, X. Cui, L. Zheng, L. SU, Y. Song, J. Shi, Oxygen vacancy-assisted hydrogen evolution reaction of the Pt/$WO_3$ electrocatalyst, J. Mater. Chem. A 7 (2019) 6285–6293

[24] N. Xue, R.J. Yu, C. Z. Yuan, X. Xie, Y.F. Jiang, H.Y. Zhou, T.Y. Cheang, A.W. Xu, In situ redox deposition of palladium nanoparticles on oxygen-deficient tungsten oxide as efficient hydrogenation catalysts, RSC Adv. 7 (2017) 2351.

[25] Y. Wang, J. Cai, M. Wu, J. Chen, W. Zhao, Y. Tian, T. Ding, J. Zhang, Z. Jiang, Z. Li, Rational construction of oxygen vacancies onto tungsten trioxide to improve visible

light photocatalytic water oxidation reaction, Appl. Catal. B 239 (2018) 398–407.

[26] C.J. Wright, Inelastic neutron scattering spectra of the hydrogen tungsten bronze $H_{0.4}WO_3$, J. Solid State Chem. 20 (1977) 89–92.

[27] W. Feng, G. Wu, G. Gao, Ordered mesoporous $WO_3$ film with outstanding gasochromic properties, J. Mater. Chem A 2 (2014) 585–590.

[28] Y.J. Xu, Y. Zhuang, X. Fu, New Insight for Enhanced Photocatalytic Activity of $TiO_2$ by Doping Carbon Nanotubes: A Case Study on Degradation of Benzene and Methyl Orange, J. Phys. Chem. C 114 (2010), 2669-2676.

[29] Y. Zhang, Z.R. Tang, X. Fu, Y.J. Xu, $TiO_2$−Graphene Nanocomposites for Gas-Phase Photocatalytic Degradation of Volatile Aromatic Pollutant: Is $TiO_2$−Graphene Truly Different from Other $TiO_2$−Carbon Composite Materials?, ACS Nano 4 (2010) 7303-7314.

[30] T. Shen, C. Jiang, C. Wang, J. Sun, X. Wang, X. Li, A $TiO_2$ modified abiotic–biotic process for the degradation of the azo dye methyl orange, RSC Adv. 5 (2015) 58704–58712.

[31] J. Fan, Y. Guo, J. Wang, M. Fan, Rapid decolorization of azo dye methyl orange in aqueous solution by nanoscale zerovalent iron particles, J. Hazard. Mater. 166 (2009) 904–910.

[32] J. Chen, J.C. Wenger, D.S, Venables, Near-Ultraviolet Absorption Cross Sections of Nitrophenols and Their Potential Influence on Tropospheric Oxidation Capacity, J. Phys. Chem. A 115 (2011) 12235–12242.

[33] J. Ran, J. Zhang, J. Yu, M. Jaroniec, S.Z. Qiao, Earth-abundant cocatalysts for semiconductor-based photocatalytic water splitting, Chem. Soc. Rev. 43 (2014) 7787–7812.

[34] W. Tongamp, Q. Zhang, F. Saito, Hydrogen generation from polyethylene by

milling and heating with Ca(OH)$_2$ and Ni(OH)$_2$, Int. J. Hydrog. Energy 33 (2008) 4097–4103.

[35] M. Senna, H. Noda, Y. Xin, H. Hasegawa, C. Takai, T. Shirai, M. Fuji, Solid-state reduction of silica nanoparticles via oxygen abstraction from SiO$_4$ units by polyolefins under mechanical stressing, RSC Adv. 8 (2018) 36338.